\documentclass[showpacs,prl,twocolumn,floatfix,amssymb]{revtex4}
\usepackage{amsmath}
\usepackage{graphicx}
\usepackage{calc}
\usepackage{bbm}
\begin{document}
\author{Wolfgang Mauerer}\email{wolfgang.mauerer@ioip.mpg.de} 
\author{Malte Avenhaus} 
\author{Wolfram Helwig} 
\author{Christine Silberhorn}
\affiliation{Max Planck Research Group, Institute of Optics, 
  Information and Photonics, Junior Research Group IQO} 
\title{How Colors Influence Numbers: Photon Statistics of Parametric 
  Downconversion} 
\pacs{42.50.Ar 89.60.Gg} 
\date{\today}
\newcommand{\unit}[1]{\ensuremath{\text{#1}}} \newcommand{\ie}{\emph{i.e.}}
\newcommand{\ket}[1]{\ensuremath{|#1\rangle}}
\newcommand{\bra}[1]{\ensuremath{\langle#1|}}
\newcommand{\ketbra}[2]{\ket{#1}\bra{#2}} \newcommand{\eg}{\emph{e.g.}}
\newcommand{\sech}{\ensuremath{\operatorname{sech}}}
\newcommand{\diag}{\ensuremath{\operatorname{diag}}}
\newcommand{\eref}[1]{(\ref{#1})}
%\renewcommand{\paragraph}[1]{}
%renewcommand{\subparagraph}[1]{}
\newcommand{\letter}{paper}

\begin{abstract}
  Parametric downconversion (PDC) is a technique of ubiquitous experimental
  significance in the production of non-classical, photon-number correlated
  twin beams.  Standard theory of PDC as a two-mode squeezing process predicts
  and homodyne measurements observe a thermal photon number distribution per
  beam. Recent experiments have obtained conflicting distributions. In this
  \letter{}, we explain the observation by an \emph{a-priori} theoretical model
  solely based on directly accessible physical quantities. We compare our
  predictions with experimental data and find excellent agreement.
\end{abstract}
\maketitle

\subparagraph{Introduction} Spectral properties of states generated by
\(\chi^{(2)}\) nonlinearities are traditionally studied using homodyne
detection. Unfortunately, this standard technique implicitly restricts the
observation to an effective single spectral mode imposed by the single local
oscillator. Avalanche photo diodes (APDs)~\cite{Achilles2003}, in contrast,
are sensitive on all modes generated by sources of current experimental
significance, and uncover richer spectral properties. This
sub-structure is currently usually neglected or only treated effectively,
although it impacts security proofs of quantum key distribution or the
validity of fundamental quantum measurements, for example.

In this \letter, we present an \emph{a-priori} theoretical explanation that
connects the spectral structure of PDC states with the photon number
distribution (PND), which is a commonly employed resource.  Recent experiments
have observed that the PND for multi-mode sources differs markedly from the
prediction of the single-mode standard model~\cite{Avenhaus2008}. Our approach
explains this behavior by decomposing the state into a set of
independent two-mode squeezers~\cite{Braunstein2005b,Wasilewski2006} akin, but
not completely identical to the Bloch-Messiah decomposition.  The PND is
inferred from the well-known properties of these independent contributions. In
contrast to previous efforts~\cite{Perina2003,Wasilewski2008}, our approach is
the first to enable, to our knowledge, the quantitative computation of photon
number statistics \emph{without} assumptions or fitting of non-physical
parameters. This is important for a wide class of experiments ranging from
fundamental to highly applied because they require a complete understanding of
the internal structure of PDC states to fully exploit their quantum features.

\subparagraph{Decomposition} 
A multi-mode type-II
downconversion process is most conveniently studied using the interaction
Hamiltonian \(\hat{H}_{\text{int}}(t) =
\int_{V}\text{d}^{3}\vec{x}\chi^{(2)}\hat{E}_{p}^{(+)}(\vec{x},t)\hat{E}_{s}^{(-)}(\vec{x},t)
\hat{E}_{i}^{(-)}(\vec{x},t) + \text{H.c.}\)~\cite{DellAnno2006}, where the
subscripts denote pump, signal, and idler, respectively, and the tensor
\(\chi^{(2)}\) represents the second-order nonlinear susceptibility. By
assuming a classical pump and a frequency-independent \(\chi^{(2)}\) in the
spectral range of interest, it can be shown~\cite{Grice1997} that with
\(\hat{H}_{I}\equiv\int_{t_{0}}^{t}\text{d}t' \hat{H}_{\text{int}}(t')\),
\begin{equation}
  \hat{H}_{I} = C\iint\text{d}\omega_{1}\text{d}\omega_{2}f(\omega_{1},
  \omega_{2})\hat{a}^{\dagger}(\omega_{1})\hat{b}^{\dagger}(\omega_{2}) + \text{H.c.},
  \label{eq:pdc_hamiltonian}
\end{equation}
where \(\hat{a}^{\dagger}(\omega_{1})\) and \(\hat{b}^{\dagger}(\omega_{2})\)
are field operators that create a monochromatic photon with frequency
\(\omega_{i}\) in the signal and idler modes a and b. \(f(\omega_{1},
\omega_{2})\) is the spectral distribution function (SDF) of the single photon
contribution, and \(C = C(\chi^{(2)}, \sqrt{I_{p}})\) is a coupling constant
that depends on the strength \(\chi^{(2)}\) of the nonlinear susceptibility
and on the pump intensity~\cite{DellAnno2006,Kolobov1999}. The time-propagated
state is computed by \(\ket{\psi} =
\mathcal{T}\exp((i\hbar)^{-1}\hat{H}_{I})\ket{\psi(t_{0})}\), where we assume
that the pulse has completely left the crystal and the interaction is
finished. Following~\cite{Grice1997}, the time-ordering \(\mathcal{T}\) can be
omitted because the Hamiltonian approximately commutes with itself at
different times and the corrections are therefore negligible.

To express \(\hat{H}_{I}\) in a more convenient form, we use the Schmidt
decomposition, uniquely defined by
\begin{equation}
  f(\omega_{1},\omega_{2}) = \sum_{n=0}^{N-1}\sqrt{\lambda_{n}}\xi_{n}^{(1)}(\omega_{1})\xi_{n}^{(2)}(\omega_{2}),
  \label{eq:schmidt_decomposition}
\end{equation}
where the Schmidt modes \(\{\xi_{n}^{(1)}(\omega_{1})\}\) and
\(\{\xi_{n}^{(2)}(\omega_{2})\}\) are two sets of orthonormal bases with
respect to the \(L^{2}\) inner product, and the Schmidt eigenvalues
\(\lambda_{n}\) are real expansion coefficients that satisfy
\(\sum_{n}\lambda_{n} =1\).  The salient feature of
Eq.~\eref{eq:schmidt_decomposition} is that only a single summation index is
required, and not two as for a regular change of basis.  The decomposition is
guaranteed to exist for a large class of systems under very general
assumptions~\cite{Parker2000}.  For simple systems that require only a few
Schmidt modes (\ie, \(N\) is small), the decomposition can be numerically
computed by solving a set of coupled integral equations~\cite{Law2000}.  For
systems that require a large \(N\), it is usually easier to perform a singular
value decomposition (SVD), see Ref.~\cite{Avenhaus2008b} and below for more
details.

We define effective single-mode field operators (sometimes also
called pseudo-boson operators) by
\begin{equation}
  \hat{A}_{n}^{\dagger} \equiv \int\text{d}\omega\xi_{n}^{(1)}(\omega)
  \hat{a}^{\dagger}(\omega),
  \label{eq:eff_ops}
\end{equation}
and similarly for \(\hat{B}_{n}\). Because the spectral distribution functions
are orthonormal, that is, \(\langle \xi_{i}, \xi_{j}\rangle = \delta_{ij}\), it is
easy to verify that the operators fulfill the canonical commutation relations
\([\hat{A}_{j}, \hat{A}_{k}^{\dagger}] = \hat{\mathbbm{1}}\delta_{jk}\) and
\([\hat{A}_{j}, \hat{A}_{k}] = 0\). More details about this notation are
provided by Ref.~\cite{Rohde2007}.

By rewriting \(\hat{H}_{I}\) in Eq.~\eref{eq:pdc_hamiltonian} using
the Schmidt decomposition~\eref{eq:schmidt_decomposition} for \(f(\omega_{1}, \omega_{2})\) and the
definition of pseudo-boson operators in Eq.~\eref{eq:eff_ops}, we obtain
\begin{equation}
  \exp\left(\frac{1}{i\hbar}\hat{H}_{I}\right) = 
  \exp\left(\frac{C}{i\hbar}\sum_{n=0}^{N-1}\sqrt{\lambda_{n}}\hat{A}^{\dagger}_{n}
      \hat{B}^{\dagger}_{n} + \text{H.c.}\right).
  \label{eq:rewritten_hamiltonian}
\end{equation}
The two-mode squeezing operator for spectral effective single modes \(A\),
\(B\) is defined by \(\hat{S}_{\text{AB}}(\eta_{n}) \equiv \exp(-\eta_{n}
\hat{A}^{\dagger}\hat{B}^{\dagger} + \eta_{n}^{*}\hat{A}\hat{B})\), where
\(\eta_{n} = C\sqrt{\lambda_{n}}/(i\hbar) \equiv r_{n} \exp{i\varphi_{n}}\) is
a complex number. Because \([\hat{A}_{j}, \hat{A}_{k}^{\dagger}] = 0\) for \(j
\neq k\), the state after the interaction is a tensor product of independent
two-mode squeezers\footnote{By re-coupling
  (neglecting unimportant phases) \(\hat{A}_{j} \rightarrow
  1/\sqrt{2}(\hat{C}_{j} - \hat{D}_{j})\), \(\hat{B}_{j} \rightarrow
  1/\sqrt{2}(\hat{D}_{j} + \hat{C}_{j})\), it follows that
  \(\hat{S}_{A_{j}B_{j}}(\eta) =
  \hat{S}_{C_{j}}(\eta)\otimes\hat{S}_{D_{j}}(\eta)\), that is, a product of
  two independent effective single-mode squeezers~\cite{Barnett1997}.  Using
  this transformation, we obtain the standard Bloch-Messiah decomposition
  (see, e.g., Refs.~\cite{Braunstein2005b,Wasilewski2006}). Owing to the
  coupling of the signal mode \(A_{j}\) with the idler mode \(B_{j}\), this
  form delivers the joint photon number distribution for signal and idler,
  \(p_{\text{joint}}(n)\). It is connected to our distribution via
  \(p_{\text{joint}}(2n) = p(n)\), and \(p_{\text{joint}}(2n+1) = 0\).  When a
  degenerate PDC process (including type-I) with \(\hat{A}_{j} = \hat{B}_{j}\)
  is considered, the decomposition in Eq.~\eref{eq:rewritten_hamiltonian}
  automatically leads to the Bloch-Messiah decomposition.}:
\begin{equation}
  \ket{\psi} = \bigotimes_{n=0}^{N-1}\hat{S}_{A_{n}B_{n}}(\eta_{n})\ket{\psi(t_{0})}.
  \label{eq:first_order}
\end{equation}
Notice that it follows from this decomposition that the SDF 
is identical for all orders of photon number contributions because
creation operators that belong to different distribution functions are never
mixed.\footnote{This property can already be inferred from
  Eq.~\ref{eq:rewritten_hamiltonian} by defining the operators
  \(\hat{K}_{n}^{(+)} \equiv \hat{A}^{\dagger}_{n}\hat{B}_{n}^{\dagger}\),
  \(\hat{K}_{n}^{(-)} \equiv \hat{A}_{n}\hat{B}_{n}\), and \(\hat{K}_{n}^{(0)}
  \equiv 1/2(\hat{A}_{n}^{\dagger}\hat{A}_{n} +
  \hat{B}_{n}^{\dagger}\hat{B}_{n})\) that share the commutation relations
  \([\hat{K}_{n}^{(0)}, \hat{K}_{n}^{(\pm)}] = \pm \hat{K}_{n}^{(\pm)}\) and
  \([\hat{K}^{(-)}_{n}, \hat{K}^{(+)}_{n}] = \hat{K}_{n}^{(0)}\) of an
  \(\mathfrak{su}(1,1)\) Lie algebra, which allows us to apply a specific
  exponential operator disentangling formula~\cite[A5.18]{Barnett1997} from
  which the desired property is readily derived.}

\subparagraph{Computing  Statistical Distributions} 
For two-mode squeezed states, the 
PND in each mode is thermal, that is,
for the state
\begin{equation}
  \ket{\psi} = \hat{S}_{AB}(\eta)\ket{00} = \sum_{n=0}^{\infty}\kappa_{n}\ket{n,n},
  \label{eq:tm_squeezed_state}
\end{equation}
the distribution is given by \(p(n) = |\kappa_{n}|^{2} =
\sech^{2}r\tanh^{2n}r\) for one output mode, that is, \(N=1\). Consequently,
the photon number distribution of the multi-mode state~\eref{eq:first_order}
is given by the convolution of the distributions of all independent
squeezers. Assume that \(p_{\xi_{k}}(n)\) denotes the PND of the \(k\)th
squeezer with spectral modes \(\xi_{k}^{(i)}\). The overall PND is then given
by
\begin{equation}
 p_{\vec{\xi}}(n) = \sum_{\Theta \in n\vdash N}\prod_{m=0}^{N-1}p_{\xi_{m}}(\Theta_{m}),
  \label{eq:tot_statistics}
\end{equation}
where \(n \vdash N\) denotes the set of all partitions of \(n\) into \(N\)
parts. The distribution \(p_{\vec{\xi}}(n)\) is consequently the
convolution of all probability distributions \(p_{\xi_{i}}(n)\).

Two special cases follow directly from Eq.~\eref{eq:tot_statistics}:
When only a single effective mode contributes (\(N = 1\)), the resulting
distribution exhibits thermal behavior. When the physical process requires a
very large number of effective modes (\(N \rightarrow \infty\)), the resulting
PND is Poissonian, because it is known that a convolution of
thermal distributions converges to a Poissonian distribution in this
limit~\cite{Mandel1995}.

Computing the convolution in Eq.~\eref{eq:tot_statistics} involves summing
over numerous contributions. This is considerably simplified by using
generating functions. For coefficients \(p(n)\), they are given by the formal
power series~\cite{Mandel1995} \(g(\zeta) = \sum_{n}p(n)\zeta^{n}\). The
individual coefficients can be recovered via \(p(n) =
\frac{1}{n!}\frac{\partial^{n}}{\partial x^{n}}\left.g(\zeta)\right|_{\zeta=0}\).  For the
thermal distribution of a two-mode squeezer, the series converges analytically
to \(g_{k}(\zeta) = \frac{\sech^{2}r_{k}}{1-\zeta \tanh^{2}r_{k}}\), where
\(r_{k}\) is the strength of the \(k\)th squeezer. The generating function for
a convolution of N thermal distributions is \(\prod_{k=0}^{N-1}g_{k}(\zeta)\),
and the resulting photon number distribution is consequently
\begin{align}
  p_{\vec{\xi}}(n) = \frac{1}{n!}\left(\frac{\partial^{n}}{\partial \zeta^{n}}
  \left.\prod_{k=0}^{N-1}g_{k}(\zeta)\right)\right|_{\zeta=0}.
  \label{eq:convolutiontats}
\end{align}

\begin{figure}[htb]
  \centering\includegraphics[width=\linewidth]%
                           {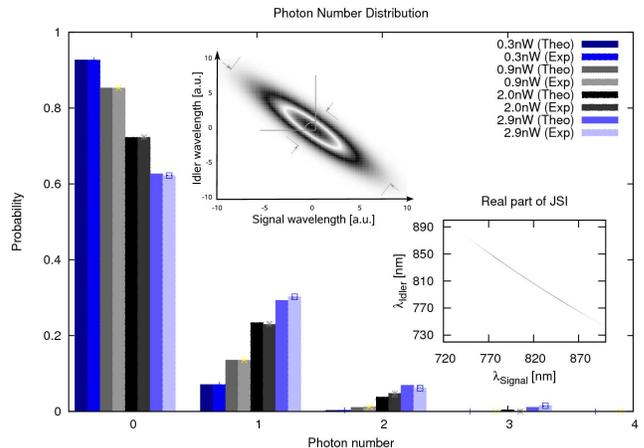}
  \caption{(Color online) Comparison between experimentally measured and
    theoretically obtained photon number distributions for a multi-mode PDC
    process at various pump strengths. The bottom inset shows the real part of
    the joint spectral intensity, while the top inset demonstrates the 
    parameterization of the analytical Gaussian approximation of the SDF.
    Loss inversion and error estimation was performed using non-negative 
    least squares optimization.}
  \label{fig:comp_with_experiment}
\end{figure}

Let us now turn our attention to an example illustrating our
considerations. Assume that the SDF is given by a two-dimensional, real-valued
Gaussian distribution (this is not a restriction because the methods also work
for complex, non-Gaussian SDFs). This approximation is commonly
used~\cite{Law2000,U'Ren2003} to provide a convenient parameterization of
type-II PDC processes. Especially, it is possible to perform an analytical
Schmidt decomposition (a similar approach is used, for instance, in
Ref.~\cite{U'Ren2003}). We use the parameters \(\sigma_{x}^{2}\) and
\(\sigma_{y}^{2}\) to specify the spectral widths of signal and idler, while
\(\theta\) denotes the rotation with respect to the \(x\) axis. This form is
illustrated in Figure~\ref{fig:comp_with_experiment}.

\begin{figure}[htb]
    \centering\includegraphics[width=\linewidth]{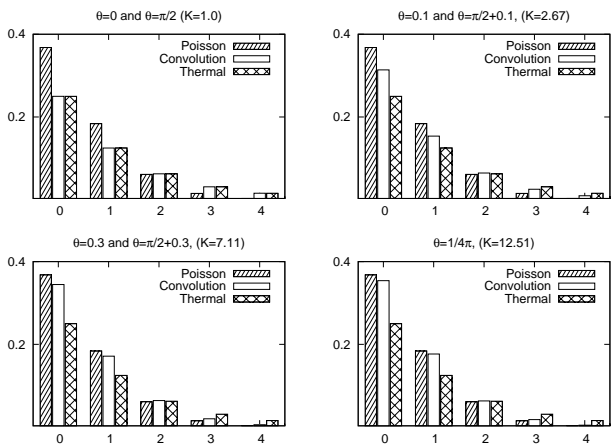}
    \caption{Photon number distribution depending on the number of effectively
      contributing modes as given by the Schmidt number \(K\) (and thus on the
      angle of the SDF) of a type-II PDC process. The \(x\) axes depict photon
      numbers, whereas the \(y\) axes show probabilities.}
    \label{fig:stat_analyt}
  \end{figure}

Let us choose \(\sigma_{x}^{2} = 25\) and \(\sigma_{y}^{2} = 1\), which are
the parameters depicted in the inset of
Figure~\ref{fig:comp_with_experiment}. The Schmidt number \(K =
1/\sum_{n}\lambda_{n}^{2}\) is computed from the eigenvalues \(\lambda_{n}\)
of the Schmidt decomposition.  It is a measure for the number of effectively
contributing spectral modes and thus of inherent spectral correlations of the
physical process~\cite{Law2000} (notice that we could have also considered an
entanglement monotone like the logarithmic negativity for this purpose).
For \(\theta = 0\), the state exhibits no spectral correlations, and a single
Schmidt mode suffices for the decomposition. By rotating the SDF from \(\theta
= 0\) to \(\theta = \pi/2\), the correlations increase to their maximal value
at \(\theta = \pi/4\), and decrease again until the SDF becomes separable for
\(\theta = \pi/2\). This implies thermal statistics for \(\theta = 0\) and
\(\theta = \pi/2\), and maximal similarity to Poissonian statistics for
\(\theta = \pi/4\). The coupling and pump intensity are, for better
comparability, chosen such that \(\bar{n} = 1\) for all PNDs.
Figure~\ref{fig:stat_analyt} illustrates the arising distributions.

\begin{figure}[htb]
  \centering\includegraphics[angle=270,width=\linewidth]{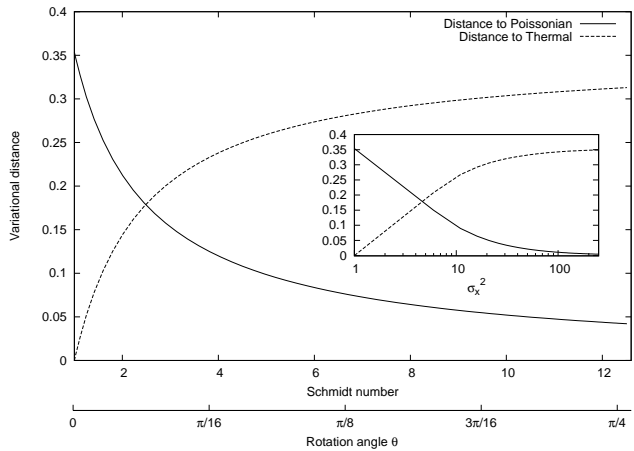}
  \caption{Solid line and dashed line show the distance between the convoluted
    photon-number distribution and Poissonian or thermal statistics,
    respectively, plotted against Schmidt number. For a single effective mode,
    the distribution is exactly thermal, but the more modes contribute, the
    closer it gets to a Poissonian distribution. The inset fixes
    \(\theta=\pi/4\) and varies \(\sigma_{x}^{2}\), which is drawn on a
    logarithmic scale.}
  \label{fig:rot_state}
\end{figure}

  To quantify the difference between convoluted and Poissonian or thermal
  distributions, we employ the variational distance defined for two
  probability distributions \(p_{1}\), \(p_{2}\) as \(\Delta_{p_{1}, p_{2}}
  \equiv \sum_{n}|p_{1}(n)-p_{2}(n)|\).  Two distributions are completely
  identical if and only if \(\Delta = 0\). Figure~\ref{fig:rot_state}
  compares the difference of the convoluted distribution to the
  above-mentioned special cases for a growing Schmidt number \(K\), that is, a
  growing number of Schmidt modes achieved by rotating the Gaussian SDF for
  \(\theta=0\) to \(\theta=\pi/4\).

Once again, we emphasize that the shift towards a Poissonian distribution is
inherent in the physical process and not caused by any experimental
imperfections.

\subparagraph{Comparison with Experimental Data} We have also performed a
comparison of experimentally measured photon number statistics with the
predictions of our theory. A photon-number resolving fiber-loop
detector~\cite{Achilles2003} in combination with highly efficient waveguides
was used to record the distribution.  The detection method is resilient
against loss and allows us to eliminate the corresponding effects when
ensemble measurements are performed. Ref.~\cite{Avenhaus2008b} shows the
experimental details of state generation, and~\cite{Avenhaus2008} describes
the measurement procedure.  Figure~\ref{fig:comp_with_experiment} compares the
experimentally observed distribution with the theoretical prediction at
various pump powers.  As is immediately obvious from the figure, they are in
excellent agreement.

To avoid the necessity of fitting any effective parameters, we have obtained
an exact numerical decomposition using SVD techniques. After discretizing 
the SDF on a grid \(M_{mn}\) of size \(1500\times 1500\), the matrix is
decomposed as \(M = U\Sigma V^{\dagger}\), where \(U\), \(V\) are unitaries
and \(\Sigma=\diag(\sqrt{\lambda_{1}}, \dots, \sqrt{\lambda_{N}})\) is a real
diagonal matrix~\cite{Golub1989}.  Extensive checks that the decomposition
converges (and also converges to the proper value) have been performed, see
Ref.~\cite{Mauerer2008b} for details.

Notice that the decomposition of the spectral distribution does \emph{not}
depend on the pump intensity, which means that the composition
\(\{\lambda_{n}\}\) of the PND is fixed for the physical process. However, the
observed \emph{mean value} of the PND does depend on the pump
intensity, and Fig.~\ref{fig:comp_with_experiment} shows a shift toward larger
mean photon numbers for larger pump intensities as expected.

\begin{figure}[htb]
  \centering\includegraphics[width=\linewidth]{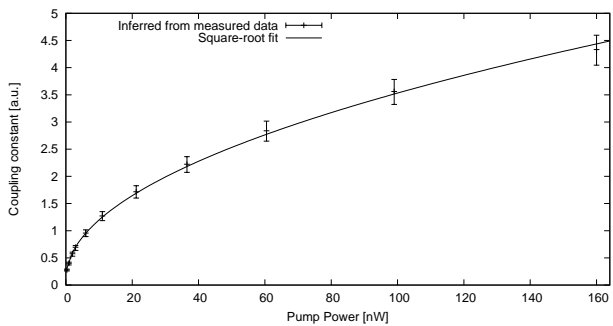}
  \caption{Relation between pump power and coupling parameter.  The expected
    square-root dependency~\cite{Kolobov1999} is correctly obtained for a
    wider range of pump intensities than can be resolved with current TMDs,
    which ensures the validity of our approach also for high powers. Notice
    that this knowledge would also allow for computing the expected mean
    photon number for a given pump power or a determination of \(\chi^{(2)}\),
    as described in~\cite{Mauerer2009}.}
  \label{fig:coupling_constant}
\end{figure}

For higher pump powers, photon-number resolved detection is not possible
anymore. To check the theory in this regime, we have used a set of mean photon
number (\(\bar{n}\)) measurements instead. The coupling constant \(C\) as
defined in Eq.~\ref{eq:pdc_hamiltonian} can be inferred from the decomposed
SDF for each \(\bar{n}\) for a given pump power by a numerical optimization
process\footnote{If \(\chi^{(2)}\) were known precisely, the optimization
  would not be necessary}. The result is shown in
Figure~\ref{fig:coupling_constant}.  Again, very good agreement between theory
and experiment is achieved.

%TODO: Show how the explanation attempts of Perina, de Riedmatten,
%and Wasilewski are contained in our considerations?

\subparagraph{Conclusions} We have shown how to decompose a
multi-mode PDC process into independent two-mode squeezers operating on
effective single modes, and how this explains why the photon number
distribution of the process can exhibit any form ranging from purely thermal
to purely Poissonian. We have underlined the validity of the theory by
comparing the predictions to an experimentally measured photon number
distribution. Additionally, we have compared theory and experiment for larger
pump powers.
\begin{appendix}
  \subparagraph{Appendix} A two-dimensional Gaussian distribution in a
  suitable parameterization is given by
  \begin{align*}
    f(x,y) &=
    \frac{1}{\sqrt{\pi\sigma_{x}\sigma_{y}}}\exp(-ax^{2}-2bxy-cy^{2}),\\
    a(\theta, \sigma_{x}, \sigma_{y}) &= \cos^2\theta/(2\sigma_{x}^{2}) +
    \sin^2\theta/(2\sigma_{y}^{2}), \\
    b(\theta, \sigma_{x}, \sigma_{y}) &=
    -\sin2\theta/(4\sigma_{x}^{2}) + \sin2\theta/(4\sigma_{y}^{2}), \\
    c(\theta,\sigma_{x}, \sigma_{y}) &= \sin^2\theta/(2\sigma_{x}^{2}) +
    \cos^2\theta/(2\sigma_{y}^{2}).
  \end{align*}

  Without getting into details of the algebra involved, we remark that by
  starting from Mehler's formula~\cite{Doetsch1930}
  \(\sum_{n=0}^{\infty}H_{n}(x)H_{n}(y)\frac{(\frac{1}{2}\gamma)^{n}}{n!} =
  \frac{1}{\sqrt{1-\gamma^{2}}} \exp\left(-\frac{\gamma^{2}x^{2}-2\gamma xy +
      \gamma^{2}y^{2}}{1-\gamma^{1}}\right)\) (\(H_{n}(x)\) denotes the
  Hermite polynomial of \(n\)th order), it is possible to bring \(f(x,y)\)
  into the form \( f(x,y) =
  \sum_{n=0}^{\infty}\sqrt{\lambda_{n}}f^{(1)}_{n}(x)f^{(2)}_{n}(y)\).  The
  coefficients \(\lambda_{n}\) are given by \(\lambda_{n} =
  \frac{2^{2n-1}}{ac} \frac{1+\gamma^{2}}
  {\sigma_{x}\sigma_{y}}\left(\frac{\gamma}{2}\right)^{2n}\) where \(\gamma =
  \frac{-2\sqrt{ac} + \sqrt{4ac - 4b^{2}}}{2b}\). Since the set
  \(\{\lambda_{n}\}\) contains all information required for our calculations,
  the exact form of \(f_{n}^{(i)}(\cdot)\) is not of interest here, but can be
  found in Ref.~\cite{Mauerer2009}.
\end{appendix}

This work was supported by the EC under the FET-Open grant agreement
CORNER, number FP7-ICT-213681.
\bibliography{literature}
\end{document}